\documentclass[prl,twocolumn,superscriptaddress,nobibnotes,citeautoscript,graphicx,amsmath,amssymb]{revtex4-1}
\usepackage{graphicx}         
\usepackage{bm}               
\usepackage{siunitx}
\usepackage{color}
\definecolor{darkblue}{rgb}{0.0,0.0,0.3}
\definecolor{goodblue}{rgb}{0.0,0.0,0.6}
\usepackage{float}
\usepackage{hyperref}
\usepackage{xfrac}
\usepackage{placeins}
\hypersetup{
	colorlinks   = true, 
	urlcolor     = darkblue, 
	linkcolor    = darkblue, 
	citecolor   = darkblue 
}

\begin{document}
\title{Spin-Orbit Protection of Induced Superconductivity in Majorana Nanowires} 

\author{Jouri D.S. Bommer}
\altaffiliation{Correspondence to \href{mailto:JouriBommer@gmail.com}{JouriBommer@gmail.com} and/or \href{mailto:HaoZhangDelft@gmail.com}{HaoZhangDelft@gmail.com} }
\affiliation{QuTech, Delft University of Technology, 2600 GA Delft, The Netherlands}
\affiliation{Kavli Institute of Nanoscience, Delft University of Technology, 2600 GA Delft, The Netherlands}

\author{Hao Zhang\footnotemark[1]}
\altaffiliation{Correspondence to \href{mailto:JouriBommer@gmail.com}{JouriBommer@gmail.com} and/or \href{mailto:HaoZhangDelft@gmail.com}{HaoZhangDelft@gmail.com} }
\affiliation{QuTech, Delft University of Technology, 2600 GA Delft, The Netherlands}
\affiliation{Kavli Institute of Nanoscience, Delft University of Technology, 2600 GA Delft, The Netherlands}
\affiliation{State Key Laboratory of Low Dimensional Quantum Physics, Department of Physics, Tsinghua University, Beijing 100084, China}

\author{\"Onder~G\"ul}
\altaffiliation{Present address: Department of Physics, Harvard University, Cambridge, MA 02138, USA}
\affiliation{QuTech, Delft University of Technology, 2600 GA Delft, The Netherlands}
\affiliation{Kavli Institute of Nanoscience, Delft University of Technology, 2600 GA Delft, The Netherlands}

\author{Bas Nijholt}
\affiliation{Kavli Institute of Nanoscience, Delft University of Technology, 2600 GA Delft, The Netherlands}

\author{Michael Wimmer}
\affiliation{QuTech, Delft University of Technology, 2600 GA Delft, The Netherlands}
\affiliation{Kavli Institute of Nanoscience, Delft University of Technology, 2600 GA Delft, The Netherlands}

\author{Filipp N. Rybakov}
\affiliation{Department of Physics, KTH-Royal Institute of Technology, SE-10691 Stockholm, Sweden}

\author{Julien Garaud}
\affiliation{Laboratoire de Math\'{e}matiques et Physique Th\'{e}orique CNRS/UMR 7350, Institut Denis Poisson FR2964, Universit\'{e} de Tours, Parc de Grandmont, 37200 Tours, France}

\author{Donjan Rodic}
\affiliation{Institut f\"{u}r Theoretische Physik, ETH Z\"{u}rich, 8093 Z\"{u}rich, Switzerland}

\author{Egor Babaev}
\affiliation{Department of Physics, KTH-Royal Institute of Technology, SE-10691 Stockholm, Sweden}

\author{Matthias Troyer}
\affiliation{Institut f\"{u}r Theoretische Physik, ETH Z\"{u}rich, 8093 Z\"{u}rich, Switzerland}
\affiliation{Microsoft Quantum, Redmond, WA 98052, USA}

\author{Diana Car}
\affiliation{Department of Applied Physics, Eindhoven University of Technology, 5600 MB Eindhoven, The Netherlands}

\author{S\'ebastien R. Plissard}
\altaffiliation{Present address: CNRS-Laboratoire d'Analyse et d'Architecture des Syst\`emes (LAAS), Universit\'e de Toulouse, 7 avenue du colonel Roche, F-31400 Toulouse, France}
\affiliation{Department of Applied Physics, Eindhoven University of Technology, 5600 MB Eindhoven, The Netherlands}

\author{Erik P.A.M. Bakkers}
\affiliation{QuTech, Delft University of Technology, 2600 GA Delft, The Netherlands}
\affiliation{Kavli Institute of Nanoscience, Delft University of Technology, 2600 GA Delft, The Netherlands}
\affiliation{Department of Applied Physics, Eindhoven University of Technology, 5600 MB Eindhoven, The Netherlands}

\author{Kenji Watanabe}
\affiliation{Advanced Materials Laboratory, National Institute for Materials Science, 1-1 Namiki, Tsukuba, 305-0044, Japan}

\author{Takashi Taniguchi}
\affiliation{Advanced Materials Laboratory, National Institute for Materials Science, 1-1 Namiki, Tsukuba, 305-0044, Japan}

\author{Leo P. Kouwenhoven}
\affiliation{QuTech, Delft University of Technology, 2600 GA Delft, The Netherlands}
\affiliation{Kavli Institute of Nanoscience, Delft University of Technology, 2600 GA Delft, The Netherlands}
\affiliation{Microsoft Station Q Delft, 2600 GA Delft, The Netherlands}

\date{\today}

\begin{abstract}
Spin-orbit interaction (SOI) plays a key role in creating Majorana zero modes in semiconductor nanowires proximity coupled to a superconductor. We track the evolution of the induced superconducting gap in InSb nanowires coupled to a NbTiN superconductor in a large range of magnetic field strengths and orientations. Based on realistic simulations of our devices, we reveal SOI with a strength of 0.15--0.35 eV\AA. Our approach identifies the direction of the spin-orbit field, which is strongly affected by the superconductor geometry and electrostatic gates.
\end{abstract}


\maketitle 

Spin-orbit interaction (SOI) is a relativistic effect that results from electrons moving (orbit) in an electric field ($E$) experiencing a magnetic field ($B_{\mathrm{SO}}$) in their moving reference frame that couples to the electron's magnetic moment (spin). SOI is an essential ingredient of various realizations of topological superconductors, which host Majorana zero modes, the building blocks of topological quantum computation \cite{Kitaev2001,Fu2008,Nayak2008}. The prime platform for topological quantum computation is based on a semiconductor nanowire coupled to a superconductor, where the proximity effect opens a superconducting energy gap in the density of states of the nanowire \cite{Lutchyn2010,Oreg2010}. In general, a magnetic field suppresses superconductivity by closing the superconducting gap due to Zeeman and orbital effects \cite{Nijholt2016}. If the nanowire has strong SOI, suppression of the superconducting gap is counteracted and a sufficiently large Zeeman energy drives the system into a topological superconducting phase, with Majorana zero modes localized at the wire ends \cite{Lutchyn2010,Oreg2010}. The main experimental effort in the last few years has focused on detecting these Majorana zero modes as a zero-bias peak in the tunneling conductance \cite{Mourik2012,Albrecht2016,Deng2016,BalMaj,QZBP,LutchynReview,AguadoReview}. However, SOI, the mechanism providing the topological protection, has been challenging to detect directly in Majorana nanowires. 
\\ \indent
The electric field that gives rise to SOI in our system mainly results from structural inversion asymmetry of the confinement potential (Rashba SOI), which depends on the work function difference at the interface between the nanowire and the superconductor and on voltages applied to nearby electrostatic gates \cite{Vuik2016,Antipov2018,Woods2018,Mikkelsen2018}. The Rashba SOI in nanowires has been investigated extensively by measuring spin-orbit related quantum effects: level repulsion of quantum dot levels \cite{Fasth2007,NadjPerge2012}, and of Andreev states \cite{DeMoor2018,Deng2016}, weak antilocalization in long \mbox{diffusive} wires \cite{Hansen2005,VanWeperen2015}, and a helical liquid signature in short quasiballistic wires \cite{JakobHelical}.
However, the SOI strength relevant to the topological protection is affected by the \mbox{presence} of the superconductor, necessitating direct observation of SOI in Majorana nanowires. Here, we reveal SOI in an InSb nanowire coupled to a NbTiN superconductor through the dependence of the superconducting gap on the magnetic field, both strength and orientation. We find that the geometry of the superconductor on the nanowire strongly modifies the direction of the spin-orbit field, which is further tunable by electrostatic gating, in line with the expected modifications of the electric field due to work function difference and electrostatic screening at the nanowire-superconductor interface.
\\ \indent
Figure 1(a) shows the device image. An InSb nanowire (blue) is covered by a NbTi/NbTiN superconducting contact (purple) and a Cr/Au normal metal contact (yellow). The barrier gate underneath the uncovered wire (red) can deplete the nanowire, locally creating a tunnel barrier. The tunneling differential conductance ($dI/dV$) resolves the induced superconducting gap, by sweeping the bias voltage ($V$) across the tunnel barrier [Fig. 1(b)]. The dashed arrow indicates the induced gap of 0.65 meV. In this device, we have recently shown ballistic transport and Majorana signatures \cite{BalMaj}.
\begin{figure}
\includegraphics[width=\columnwidth]{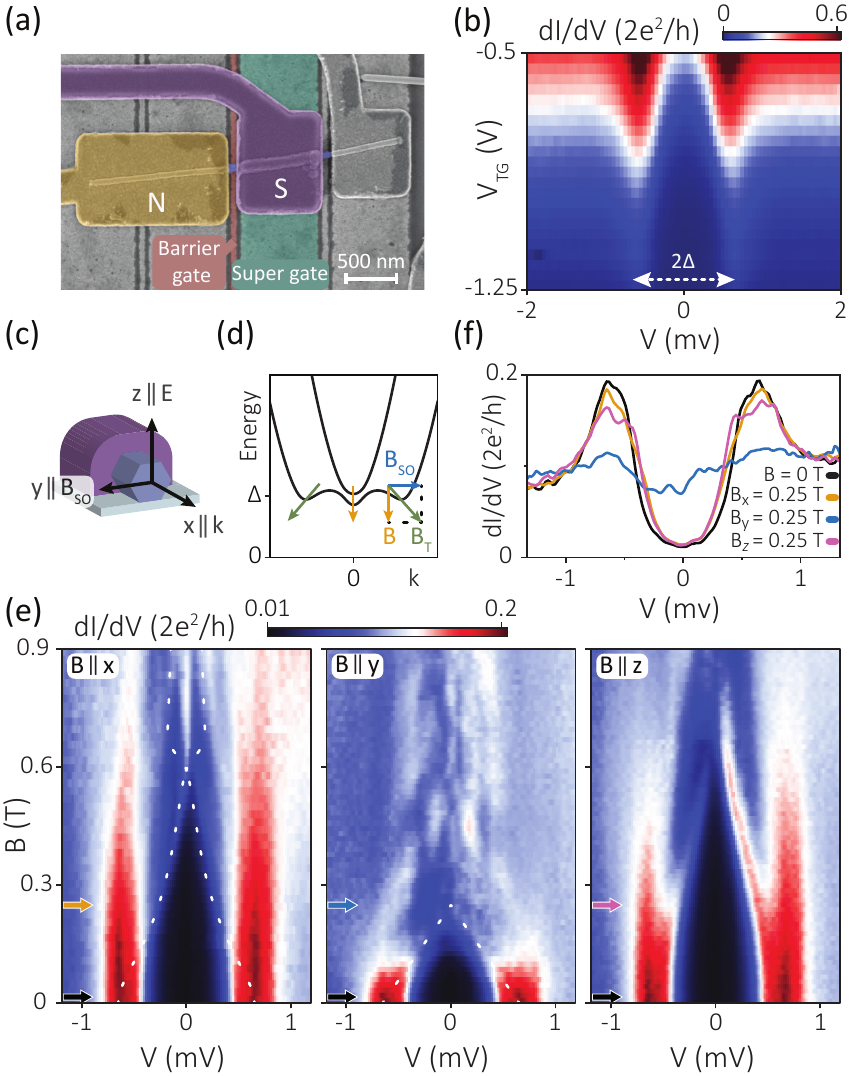}
\caption{\label{fig1}
	(a) False-color scanning electron micrograph of Majorana nanowire device $A$. An InSb nanowire (blue) is contacted by a normal metal contact ($N$, yellow) and a NbTiN superconducting contact ($S$, purple). The additional contact (gray) is kept floating. The nanowire is isolated from the barrier gate (red) and the super gate (green) by $\sim$ 30 nm thick boron nitride. (b) Differential conductance $dI/dV$ as a function of bias voltage $V$ and barrier gate voltage $V_{\mathrm{barrier}}$ at $B$ $=$ 0 T. (c) Schematic of the nanowire device and definition of the axes. (d) Band diagram of a Majorana nanowire at an externally applied magnetic field $B$ perpendicular to the spin-orbit field $B_{\mathrm{SO}}$. The arrows indicate the total magnetic field $B_T$ $=$ $B$ + $B_{\mathrm{SO}}$ along which the spin eigenstates are directed. At $k$ $=$ 0 the spin always aligns with $B$. At increasing $k$, $B_{\mathrm{SO}}$ increases, tilting the spin more towards $B_{\mathrm{SO}}$. (e) $dI/dV$ as a function of $V$ at $B$ along $x$, $y$, $z$ (left, middle, right) for super gate voltage $V_{\mathrm{SG}}$ $=$ 0 V. The white dashed lines indicate a fit to the gap closing corresponding to $\alpha$ $=$ 0.15 $\pm$ 0.05 eV\AA. (f) Horizontal line cuts of (e) at $B$ indicated by the colored arrows in (e).
}
\end{figure}
\\ \indent
The magnetic field ($B$) dependence of the induced gap of device $A$, with $B$ along three different directions, is shown in Fig. 1(e). The coordinate system is illustrated in Fig. 1(c). The $x$ axis is along the nanowire, parallel to the electron momentum ($k$). The $z$ axis is perpendicular to the substrate and coincides with the electric field ($E$) direction due to the spatial symmetry of the device and the bottom gate. Since the Rashba spin-orbit field ($B_{\mathrm{SO}}$ $\propto$ $E$ $\times$ $k$) is perpendicular to both $k$ and $E$, it points along the $y$ axis. When $B$ is aligned with $x$ or $z$ [left and right panels in Fig. 1(e)], both perpendicular to $B_{\mathrm{SO}}$, the gap closes \mbox{slowly} (at around 0.6 T), followed by the emergence of a zero-bias peak possibly characteristic of a Majorana zero mode when $B$ is along the nanowire, although we emphasize that a conjecture of Majorana zero modes is not essential for the purposes of this Letter. On the contrary, when $B$ is aligned with the $y$ axis (middle panel), parallel to $B_{\mathrm{SO}}$, the gap closes much faster (at around 0.25 T). Figure 1f shows the line cuts at $|B|$ $=$ 0.25 T along the three axes: for $B$ $\perp$ $B_{\mathrm{SO}}$, the gap is almost the same as when $B$ $=$ 0 T, while the gap is closed for $B$ $\parallel$ $B_{\mathrm{SO}}$. This observation matches the predictions of the Majorana nanowire model, as illustrated in Fig. 1(d): when $B$ $\perp$ $B_{\mathrm{SO}}$, SOI counteracts the Zeeman-induced gap closing by rotating the spin eigenstate towards $B_{\mathrm{SO}}$, which reduces the component of the Zeeman field along the direction of the spin eigenstate. In contrast, when $B$ $\parallel$ $B_{\mathrm{SO}}$, the spin eigenstate is always parallel to $B$, which prevents spin-orbit protection and results in a fast gap closing \cite{Osca2014,Rex2014}. This pronounced anisotropy of the gap closing with respect to different $B$ directions is universally observed in over ten devices (four shown in this Letter) for all gate settings 
\footnote{See Supplemental Material, which includes Refs. 
 \cite{Car2014,Flohr2011,Suyatin2007,HardGap,Liu2017,Danon2017,Hofstader1976,Gropp1996,Du1999,Prada2012,Pientka2012,Stanescu2012}, for experimental details, theoretical details, and additional experimental data}, which is a direct consequence of SOI in Majorana nanowires.
\nocite{Car2014,Flohr2011,Suyatin2007,HardGap,Liu2017,Danon2017,Hofstader1976,Gropp1996,Du1999,Prada2012,Pientka2012,Stanescu2012}
\\ \indent
Before we discuss the SOI in more detail, we rule out alternative mechanisms for the anisotropy which can originate in the bulk superconductor, or the InSb nanowire. First, an anisotropic magnetic field-induced closing of the bulk superconducting gap is excluded for the fields we apply, which are far below the critical field of NbTiN ($>$9 T) \cite{DavidOneMin}. We note that this is different from aluminium films \cite{Chang2015,Deng2016,Gazibegovic2017,QZBP}, where a small magnetic field ($<$0.3 T) perpendicular to the film completely suppresses superconductivity, making them unsuitable to reveal SOI from an anisotropic gap closing. Next, we consider Meissner screening currents in NbTiN that can cause deviations in the magnetic field in the nanowire. Our Ginzburg-Landau simulations show that the field corrections due to Meissner screening are negligible \cite{Note1}, since the dimensions of the NbTiN film ($<$1 $\mu$m) are comparable to the penetration depth ($\sim$290 nm). The simulations also show that vortex formation is most favorable along the $z$ axis \cite{Note1}, which implies that the observed anisotropic gap closing is not caused by gap suppression due to vortices near the nanowire \cite{Takei2013}, since we do not observe the fastest gap closing along $z$ [Fig. 1(f)]. Finally, in the InSb nanowire, the Zeeman $g$ factor can become anisotropic due to quantum confinement \cite{NadjPerge2012,Pryor2006,Qu2016}. However, our nanowire geometry leads to confinement in both the $y$ and $z$ directions, implying similar gap closing along $y$ and $z$, inconsistent with our observations [Fig. 1(e)].
\\ \indent
Having excluded the above mechanisms, we are now left with three effects: spin splitting of the electron \mbox{states} in magnetic fields with the Land\'e $g$ factor (Zeeman eff\mbox{ect)}, the orbital effect of the magnetic field representing the Lorentz force acting on traveling electrons, and SOI. To investigate the role of these effects, we use a theoretical three-dimensional Majorana nanowire model defined by the Hamiltonian \cite{Lutchyn2010,Oreg2010,Nijholt2016}:
\begin{equation*}
\begin{split}
H = &\left(\frac{\mathbf{p}^2}{2m^*}-\mu+V(y,z)\right) \tau_z + \frac{\alpha}{\hbar} \boldsymbol{\sigma} \cdot \mathbf{(\hat{E}\times p)} \tau_z\\
&+ \frac{1}{2}g\mu_B\mathbf{B\cdot}\boldsymbol{\sigma}+\Delta_0 \tau_x
\end{split}
\end{equation*}
Here, the first term represents the kinetic and potential energy, with $\mu$ the chemical potential measured from the middle of the helical gap and $V(y,z)$ $=$ $\frac{\Delta V_G}{R}[0,y,z]$ $\cdot$ $\mathbf{\hat{E}}$ is the electrostatic potential in the wire, whose magnitude is parametrized by $\Delta V_G$, with $\mathbf{\hat{E}}$ the direction of the electric field and $R$ the wire radius. The orbital effect enters the Hamiltonian via the vector potential $\mathbf{A}$ in the canonical momentum: $\mathbf{p}$ $=$ $-i\hbar \nabla$ $+$ $e\mathbf{A}$. Here, $e$ is the electron charge, $\hbar$ is Plank's constant, and $m^*$ $=$ 0.015 $m_e$ is the effective mass with $m_e$ the electron mass. The second term represents Rashba SOI characterized by a SOI strength $\alpha$, which we set to 0.2 eV\AA\ to find qualitative agreement with the measurements. The third term is the Zeeman term, with an isotropic $g$ factor set to 50 and $\mu_B$ is the Bohr magneton. The last term accounts for the superconducting proximity effect, which we implement in the weak coupling approximation \cite{Nijholt2016}. The Pauli matrices $\tau$ and $\sigma$ act in the particle-hole and spin space respectively. We perform numerical simulations of this Hamiltonian on a 3D lattice in a realistic nanowire geometry using the \textsc{kwant} code \cite{Groth2014}. We note that recent theory work shows that the anisotropy is unaffected by additional factors such as the wire length, temperature, and strong coupling to the superconductor \cite{Liu2019}. Additional details are provided in the Supplemental Material \cite{Note1}.
\begin{figure}
\includegraphics[width=\columnwidth]{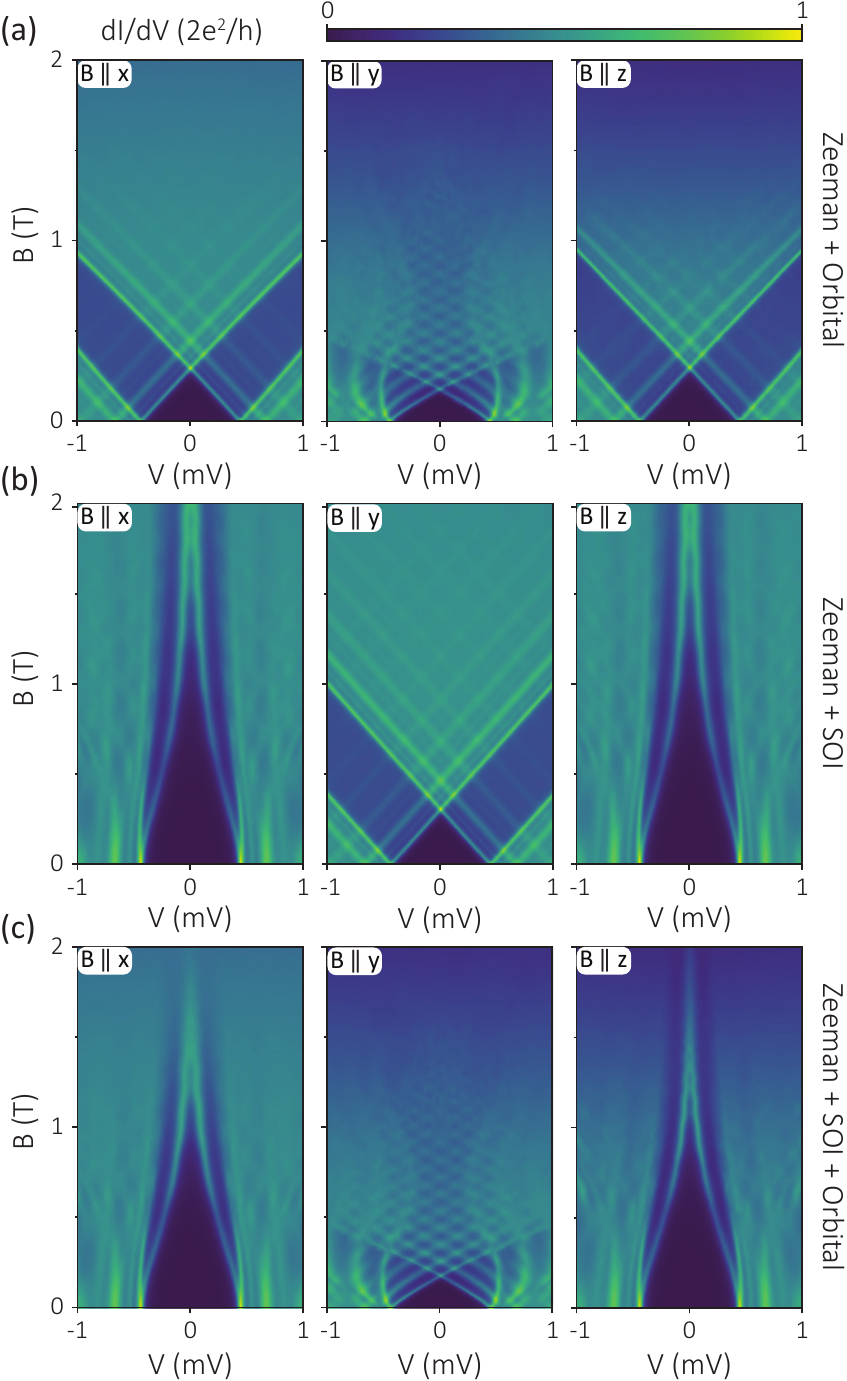}
\caption{\label{fig2}
	(a) Numerical simulations of $dI/dV$ as a function of $V$ and $B$, including the Zeeman and the orbital (Lorentz) effect of the magnetic field. (b) Same as (a), but including Zeeman and SOI instead of the orbital effect, reproducing the anisotropy in Fig. 1(e). (c) Same as (b), but including the Zeeman, SOI and orbital effect. The parameters used in (a)-(c) are $\mu$ $=$ 5.6 meV and $\Delta V_G$ $=$ -8 meV.
}
\end{figure}
\\ \indent
We identify which effects explain the observed anisotropic gap closing behavior by including them separately in our simulations. Figure 2(a) shows the magnetic field dependence of the gap without SOI (setting $\alpha$ $=$ 0 in the Hamiltonian). In contrast to Fig. 1(e) the gap closes around 0.3 T for all three directions, reflecting the dominant contribution of the Zeeman effect. In Fig. 2(b), we turn on the SOI, and turn off the orbital effect by setting the magnetic vector potential $\mathbf{A}$ $=$ 0, which qualitatively reproduces the anisotropic behavior between the $y$ axis and the $x$ and $z$-axes. We have explored other combinations of parameters and find that the experimental results of Fig. 1(e) can only be reproduced by including SOI. We note that adding the orbital effect in Fig. 2(c) shifts the gap closing to a field almost twice as small for $B$ $\parallel$ $y$, which explains why we observe a gap closing for $B$ $\parallel$ $y$ at around 0.25 T, far below 0.45 T, the critical field expected when only the Zeeman effect with $g$ $=$ 50 suppresses the gap. By fitting the curvature of the gap closing \cite{VanHeck,Pan2018} along $x$ [white dashed line in Fig. 1(e)] we estimate a range of the SOI strength $\alpha$ of 0.15 -- 0.35 eV\AA\ from devices $A$-$D$ (for fitting details and fits to additional devices, see Supplemental Material \cite{Note1}). This SOI strength is in \mbox{agreement} with the values extracted from level repulsion of Andreev states \cite{Stanescu2013,DeMoor2018} in an additional device $E$ \cite{Note1}. \mbox{Since} $\alpha$ depends on the electric field in the wire, we expect the observed variation in the SOI strength of devices to be caused by differences in the applied gate voltages and wire diameter. Recently, the level repulsion of Andreev states in InSb nanowires covered with epitaxial aluminium has shown a SOI strength of approximately 0.1 eV\AA \cite{DeMoor2018}, slightly lower than we find for NbTiN covered nanowires, most likely due to strong coupling to the aluminium superconductor, leading to stronger renormalization of the InSb material parameters \cite{Stanescu2011,Cole2015,Antipov2018,Woods2018,Mikkelsen2018,Reeg2018}.
\begin{figure}
\includegraphics[width=\columnwidth]{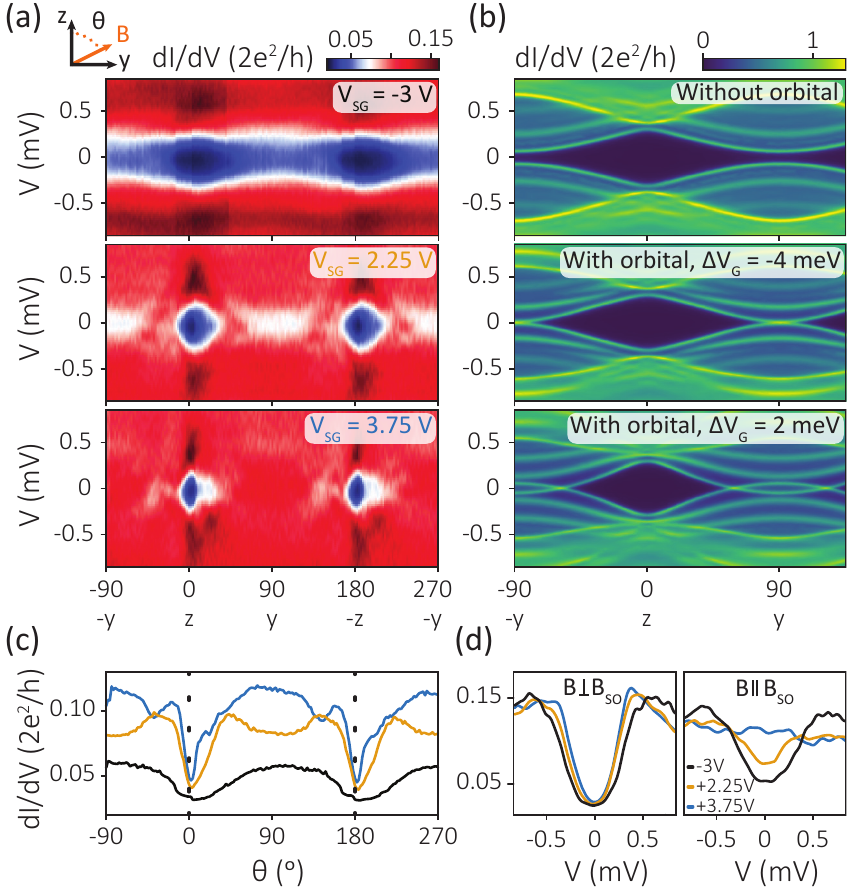}
\caption{\label{fig3}
	(a) Measured $dI/dV$ as a function of $V$ upon rotation of $B$ at 0.3 T over angles $\Theta$ between $z$ and $y$ in device $B$ (see Fig. S5 \cite{Note1} for the same behavior in device $A$). The voltage $V_{\mathrm{SG}}$ on the super gate (see insets) is varied in the three panels.
	(b) Simulated $dI/dV$ as a function of $\Theta$ and $V$ at 0.25 T. The top panel includes the Zeeman effect and SOI. The middle and bottom panels additionally include the orbital effect at two values of the potential difference $\Delta V_G$ between the top and middle of the wire. 
	(c) Horizontal line cuts of (a) averaged over $|V|$ $<$ 0.2 V at $V_{\mathrm{SG}}$ $=$ -3,  2.25, and 3.75 V (black, orange, blue). Dashed lines indicate the $z$ axis ($\Theta$ $=$ \ang{0}). (d) Vertical line cuts of (a) at $\Theta$ $=$ \ang{0} (left) and $\Theta$ $=$ \ang{90} (right).
}
\end{figure}
\\ \indent
To resolve the direction of the spin-orbit field, we fix the $B$ amplitude and continuously rotate the $B$ direction, parametrized by the angle $\Theta$ in the $zy$ plane [inset Fig. 3(a)]. Figure 3(a) shows the dependence of the gap on $\Theta$, where we adjust the electric field strength in the nanowire with a voltage $V_{\mathrm{SG}}$ on the super gate (SG) underneath the superconductor [green in Fig. 1(a)]. We define the angle at which the gap is hardest as $\Theta_{\mathrm{max}}$ and find $\Theta_{\mathrm{max}}$ $=$ 3 $\pm$ \ang{2} ($z$ axis) for all $V_{\mathrm{SG}}$ and in multiple devices (Fig. 3 and Fig. S5 \cite{Note1}) (error due to uncertainty in the extraction procedure). This is illustrated in Fig. 3(c), which shows horizontal line cuts for subgap bias. The largest gap for a given $B$ amplitude is expected for $B$ $\perp$ $B_{\mathrm{SO}}$, indicating that $B_{\mathrm{SO}}$ $\parallel$ $y$, in agreement with the $E$-field direction dictated by the device geometry.

\begin{figure}[b]
\includegraphics[width=\columnwidth]{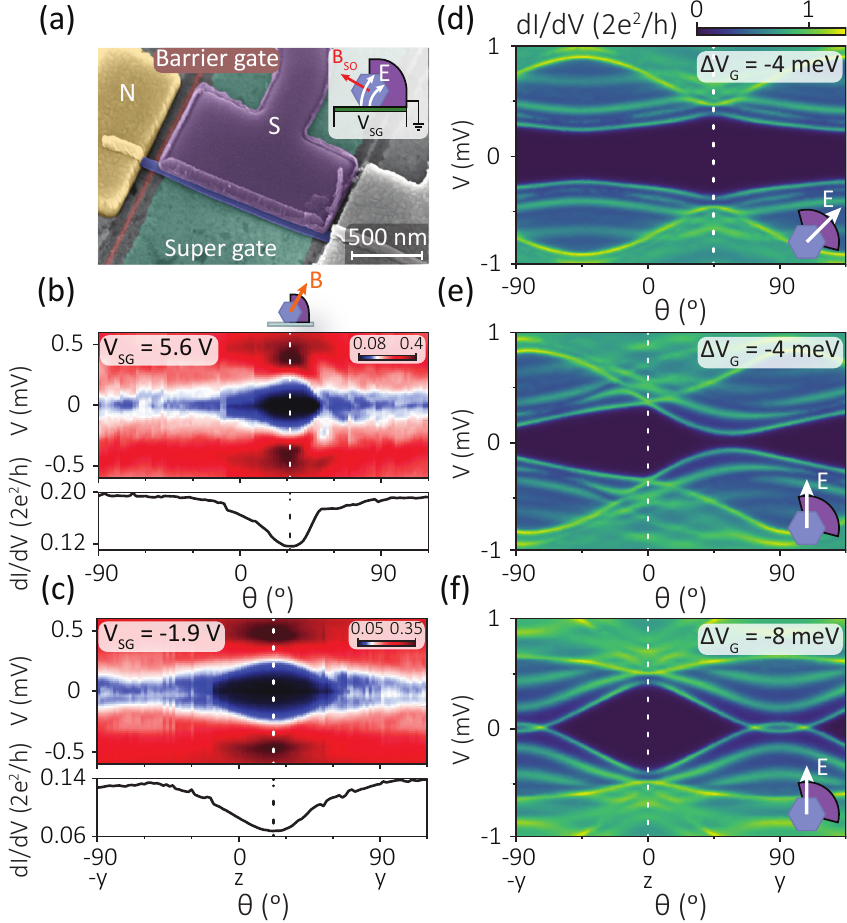}
\caption{\label{fig4}
	(a) Tilted view electron micrograph of Majorana nanowire device $E$, which is partially covered with NbTiN. In this device, the electric field $E$ (and the associated spin-orbit field $B_{\mathrm{SO}}$) can rotate away from the $z$ axis ($y$ axis), as illustrated in the inset. (b) Measured $dI/dV$ as a function of $V$ and angle $\Theta$ in the $zy$ plane at $|B|$ $=$ 75 mT and $V_{\mathrm{SG}}$ $=$ 5.6 V, with a horizontal line cut averaged over $|V|$ $<$ 0.25 mV in the lower panel. The gap is maximum at $\Theta_{\mathrm{max}}$ $=$ \ang{32} as indicated by the dashed line. (c) Same as (b), but at $V_{\mathrm{SG}}$ $=$ -1.9 V and $|B|$ $=$ 0.15 T. $\Theta_{\mathrm{max}}$ is gate tuned to \ang{22}. (d)-(f) Simulated $dI/dV$ at 0.25 T at various $\Delta V_G$ (see inset) with the superconductor rotated to the side by \ang{45} and including the Zeeman effect, SOI, and the orbital effect. The illustrations in the insets indicate the direction of $E$, which is rotated by \ang{45} from $z$ in (d).
}
\end{figure}

Now, we check whether the orbital effect changes $\Theta_{\mathrm{max}}$. The simulations in Fig. 3(b) show the effect of magnetic field rotation on the gap with $B_{\mathrm{SO}}$ $\parallel$ $y$, confirming that $\Theta_{\mathrm{max}}$ is, indeed, always given by the direction perpendicular to $B_{\mathrm{SO}}$, i.e. $\Theta_{\mathrm{max}}$ $=$ \ang{0}. Comparing the top panel (without the orbital effect) with the middle panel (with the orbital effect), we conclude that the orbital effect does not affect $\Theta_{\mathrm{max}}$. This conclusion also holds when we vary the potential difference $\Delta V_G$ between the middle and outer of the wire (corresponding to $V_{\mathrm{SG}}$) in the middle panel and bottom panel. We note that, at $\Delta V_G$ $=$ 2 meV (bottom panel) the wave function is moved towards the bottom of the nanowire, which increases the strength of the orbital effect by breaking the reflection symmetry about the $z$ axis, as evidenced by the longer angle range over which the gap is closed compared to $\Delta V_G$ $=$ -4 meV (middle panel). Experimentally, we also observe this in Fig. 3(a), with line cuts in Fig. 3(c), where the gap is closed over a significantly longer angle range with increasing $V_{\mathrm{SG}}$. We note that we use small values of $\Delta V_G$ in the simulations, because we expect a weak gate response due to effective electrostatic screening by the superconductor, which covers five of the six nanowire facets \cite{BalSc}.
\\ \indent
Finally, we turn to a second type of device in which the superconducting film only partially covers the nanowire facets [Fig. 4(a)]. This partial superconductor coverage can modify the orientation of $B_{\mathrm{SO}}$ by changing the associated electric field direction \cite{Vuik2016}, as sketched in the inset of Fig. 4(a). The electric field in the wire has two main origins. The first one originates from the work function difference between the superconductor and nanowire, which leads to charge redistribution. The resulting electric field is expected to rotate away from the $z$ axis due to the partial superconductor coverage which breaks the spatial symmetry. In Fig. 4(b) we rotate $B$ in the $zy$ plane, perpendicular to the nanowire axis, and find that $\Theta_{\mathrm{max}}$ is, indeed, no longer at zero, but at 32 $\pm$ \ang{2}. The second contribution to the electric field arises from the applied $V_{\mathrm{SG}}$ and the electrostatic screening due to the grounded superconductor. Changing $V_{\mathrm{SG}}$ should, therefore, rotate the electric field for partial coverage. Indeed, we find that $\Theta_{\mathrm{max}}$ shifts by \ang{10} by adjusting $V_{\mathrm{SG}}$ by 7.5 V [Fig. 4(c)]. Field rotation at intermediate $V_{\mathrm{SG}}$ and magnetic field sweeps confirming the change of $\Theta_{\mathrm{max}}$ are shown in the Supplemental Material \cite{Note1}. Our theory simulations confirm that $\Theta_{\mathrm{max}}$ is still given by the direction orthogonal to $B_{\mathrm{SO}}$ when the electric field is not necessarily along a \mbox{spatial} symmetry axis of the partially covered device [Fig. 4(d) and 4(e)]. While the orbital effect does not change $\Theta_{\mathrm{max}}$ [Fig. 4(e) and 4(f)], it can induce asymmetry in the energy spectrum around $\Theta_{\mathrm{max}}$ resulting from wave function asymmetry when the electric field is not along the mirror plane of the device [Fig. 4(b) and Fig. 4(e)]. The significance of the orbital effect in our devices underlines the importance of including it in realistic simulations of Majorana nanowires.
\\ \indent
In conclusion, the observed gap closing anisotropy for different magnetic field orientations demonstrates SOI in our Majorana nanowires, a necessary condition to create Majorana zero modes. Our experiments reveal that SOI is strongly affected by the work function difference at the nanowire-superconductor interface and the geometry of the superconductor, while electrostatic gating provides tunability of SOI.

\begin{acknowledgments}
We thank O.W.B. Benningshof, A. Geresdi, S. Goswami, M.W.A. de Moor, M. Quintero-P\'{e}rez and P. Ro\.{z}ek for valuable feedback and assistance. This work has been supported by the Netherlands Organization for Scientific Research (NWO), Foundation for Fundamental Research on Matter (FOM), European Research Council (ERC) and Microsoft Corporation Station Q. The work of F.N.R. and E.B. was supported by the Swedish Research Council Grant No. 642-2013-7837 and by G\"{o}ran Gustafsson Foundation for Research in Natural Sciences and Medicine.

J.D.S.B., H.Z. and \"{O}.G. contributed equally to this work.
\end{acknowledgments}

%

\end{document}